\documentstyle[preprint,aps]{revtex}
\begin{document}
\draft
\title{\bf{\LARGE{ The role of power law nonlinearity in the discrete 
nonlinear Schr\"{o}dinger equation on the formation of 
stationary localized states in the Cayley tree}}}
\author{K. Kundu and B. C. Gupta}
\address{Institute of physics, Bhubaneswar - 751 005, India}
\maketitle
\begin{abstract}
We study the formation of stationary localized states using the discrete 
nonlinear Schr\"{o}dinger equation in a Cayley tree with connectivity
$K$. Two cases, namely, a dimeric power law nonlinear impurity and a fully 
nonlinear system are considered. We introduce a transformation which reduces the 
Cayley tree into an one dimensional chain with a bond defect. The hopping 
matrix element between the impurity sites is reduced by $1/\surd K$. The 
transformed system is also shown to yield tight binding Green's function 
of the Cayley tree. The dimeric ansatz is used to find the reduced 
Hamiltonian of the system. Stationary localized  states are found from the 
fixed point equations of the Hamiltonian of the reduced dynamical system. 
We discuss the existence of different kinds of localized states. We have also 
analyzed the formation of localized states in one dimensional system with a 
bond defect and nonlinearity which does not correspond to a Cayley tree. 
Stability of the states is discussed and stability diagram is presented for 
few cases. In all cases the total phase diagram for localized states have been 
presented.
\end{abstract}

\narrowtext
\section{Introduction}

One well studied nonlinear equation in condensed matter physics and
optics is the discrete nonlinear Schr$\ddot{\rm o}$dinger 
equation (DNLSE). The DNLSE is a 
nonintegrable standard discretization of the integrable nonlinear 
Schr\"{o}dinger equation \cite{kiv}. The DNLSE in one dimension in its 
general form is a set of $n$ coupled nonlinear differential equations.
\begin{eqnarray}
i \frac{dC_{m}}{dt} &=& - \chi_{m} f_{m}(\mid C_{m} \mid) C_{m} 
+ V_{m,m+1} C_{m+1} 
+ V_{m,m-1} C_{m-1} \nonumber \\
{\rm where}~~~ V_{m,m+1} &=& V_{m+1,m}^{\star};~ {\rm and}
~m=1,2,3,...n.
\end{eqnarray}
In Eq.(1) the nonlinearity appears through functions $f_{m}(\mid 
C_m \mid)$ and $\chi_{m}$ is the nonlinearity parameter associated 
with the $m$-th grid point. Since, $\sum_{m} \mid C_m \mid^2$ is made 
unity by choosing appropriate initial conditions, $\mid C_m \mid^2$ 
can be interpreted as the probability of finding a particle at the 
$m$-th grid point. The analytical solutions of Eq.(1) in general are 
not known. Numerous works, both analytical as well as numerical, on 
the DNLSE have been reported \cite{kiv,1,2,3,4,5,6,7,8,9,10,11,12,13}.
As for its application particularly in condensed matter physics, we
cite among others the exciton propagation in Holstein molecular
crystal chain \cite{1}. In general, the exciton propagation in quasi one 
dimensional systems \cite{14}  having short range electron phonon 
interaction can
be adequately modeled by the DNLSE. Other examples include nonlinear 
optical responses in superlattices formed by dielectric or magnetic 
slabs \cite{15}  and the mean field theory of a periodic array of 
twinning planes 
in the high $T_c$ superconductors \cite{16}. 

One important feature of the DNLSE is that this can yield stationary 
localized ( SL ) states. These SL states might play a relevant role 
in the nonlinear DNA dynamics \cite{17} and also in the energy localization in 
nonlinear lattices \cite{18}. It has been shown that the presence of a 
nonlinear impurity can produce SL states in one, two and three 
dimension \cite{tsi1,tsie,yiu,hui,bik1,wein,acev,bik2,bik3}.  
The formation of SL states due 
to the presence of a single and a dimeric nonlinear impurity in 
few linear hosts has been studied in details \cite{bik1}. 
The same problem is also studied starting from an appropriate 
Hamiltonian \cite{bik2}. The fixed point of the Hamiltonian 
\cite{wein,acev,bik2,bik3} 
which generates the appropriate DNLSE can also produce the correct
equations governing the formation of SL states. We further note that 
the appropriate ansatz for the dimer problem has been obtained in our 
earlier analysis \cite{bik1}. Furthermore, the formation of inter site peaked 
and dipped stationary localized states has been studied using the 
dimeric ansatz. The effect of one nonlinear impurity as well as a dimeric
impurity in an otherwise perfect nonlinear chain on the formation of 
SL states has also been studied.\cite{bik3} The other important aspect to be 
considered is the effect of connectivity on the formation of SL states. 
The simplest way to study this is to consider the Cayley tree.
The effect of one power law nonlinear impurity in an otherwise perfect 
Cayley tree has already been considered.\cite{bik1} So, we plan here to study 
the effect of a dimeric nonlinear impurity in an otherwise perfect
Cayley tree. We further consider the fully nonlinear Cayley tree. 
For this purpose a transformation is devised to map the system 
to an effective one dimensional chain with a bond defect. Consequently,
from the symmetry consideration, the dimeric ansatz is found to be 
suitable for this study. The appropriate form of the monomeric ansatz is
also found for this case to study the formation of on-site peaked solitons.

The organization of the paper is as follows.   In sec. II we introduce 
a transformation which reduces the Cayley tree into a one dimensional 
system with a bond defect. This transformation has also been checked
through Green's function analysis in Appendix I. SL states in the Cayley 
tree in presence
of a dimeric nonlinear impurity is discussed in sec. III. In sec. IV,
fully nonlinear Cayley tree is discussed. Finally in Sec. V we summarize
our investigations.

\section{Transformation of Cayley tree to one dimensional system}

The structure of a Cayley tree with connectivity, $K=Z-1$ is shown
in Fig.(1). $Z$ is the coordination number. We pick up a connection 
and its two ends are numbered 0 and 1 respectively without any loss 
of generality. Furthermore, all points in a given generation lie 
in a shell. Shells are number by $n$ and $n \in Z$ as shown in Fig.(1).
In a perfect Cayley tree the number of points in the $n^{th}$ 
shell is $K^{n-1}$ if $n \ge 1$ and $K^{|n|}$ if $n \le 0$. We 
further note that for a perfect Cayley tree all points in a given 
shell have identical neighborhood. 

We consider now the motion of a particle on a Cayley tree with 
connectivity, $K$. In the tight binding formalism with nearest
neighbor hopping only equations governing the motion of the 
particle are 
\begin{eqnarray}
i \frac{d{\widetilde C_n}}{d{t}}&=&K \widetilde C_{n+1} + 
\widetilde C_{n-1} + \widetilde \epsilon_n \widetilde C_n , ~~~~~~~~~~~n > 1 
\nonumber \\
i \frac{d{\widetilde C_n}}{d{t}}&=&K \widetilde C_{-|n|-1} + 
\widetilde C_{-|n|+1} + \widetilde \epsilon_n \widetilde C_n , ~~~~~n < 0 
\nonumber \\
i \frac{d{\widetilde C_1}}{d{t}}&=&K \widetilde C_{2} + 
\widetilde C_{0} + \widetilde \epsilon_1 \widetilde C_1 \nonumber \\ 
i \frac{d{\widetilde C_0}}{d{t}}&=&K \widetilde C_{-1} + 
\widetilde C_{1} + \widetilde \epsilon_0 \widetilde C_0 
\end{eqnarray}
Here $\widetilde C_n$ denotes the probability amplitude at any point
in the $n^{th}$ shell and all points in the $n^{th}$ cell have the same 
probability amplitude because of identical neighborhood. The nearest 
neighbor hopping matrix, $V$ has been 
taken to be unity without any loss of generality. It is also assumed that
all points in a given shell arising due to a specific organization have 
same site energy. We, however, note that in our subsequent work with the 
DNLSE this assumption will be automatically satisfied. The normalization 
condition for the site amplitudes gives
\begin{equation}
\sum_{-\infty}^0 K^{|n|} |\widetilde C_n|^2 + \frac{1}{K} \sum_{n=1}^\infty
K^n |\widetilde C_n|^2 = 1.
\end{equation}
We now make the following transformations.
(i) $\tau = \sqrt K t$, (ii) $\epsilon_n = \widetilde \epsilon_n / \sqrt K$,
(iii) $\widetilde C_n = K^{-(n-1)/2} C_n$,~~for $n \ge 1$ and (iv) $\widetilde
C_{-|n|} = K^{-|n|/2} C_n$~~ for $n \le 0$.
After substituting these transformations in Eq.(2) we finally obtain
\begin{eqnarray}
i \frac{dC_n}{d\tau}&=&C_{n+1} + C_{n-1} + \epsilon_n C_n, ~~{\rm for}~~ n >1 
~~{\rm and}~~ n < 0. \nonumber \\
i \frac{dC_1}{d\tau}&=&C_{2} + \frac{1}{\sqrt K}C_{0} + \epsilon_1 C_1, 
\nonumber \\ 
i \frac{dC_0}{d\tau}&=&C_{-1} + \frac{1}{\sqrt K}C_{1} + \epsilon_0 C_0. 
\end{eqnarray}
Furthermore from Eq.(3) normalization condition reduces to
$\sum_{-\infty}^{\infty} |C_n|^2$ = 1. So,
the motion of a particle on a Cayley tree is mapped to that on a one 
dimensional chain. However, in this chain the nearest neighbor hopping 
matrix element between the zeroth and first site is reduced from unity 
to $1/\sqrt K$. In appendix I we show that the Green's function 
$G_{0,0}(E)$ calculated from Eq.(4) will yield the $\widetilde G_{0,0}
(\widetilde E = E\sqrt K)$ of a Cayley tree with connectivity $K$. 

Since we are interested in the DNLSE with general power law nonlinear 
impurity, in our case $\epsilon_n = \widetilde 
\epsilon_n/\sqrt K = \chi_n K^{-(n-1)\sigma /2} K^{-1/2} 
|C_n|^\sigma$ for $n \ge 1$ and $\epsilon_{-|n|} = \widetilde \epsilon_{-|n|}
/\sqrt K = \chi_{-|n|} K^{-|n|\sigma /2} K^{-1/2} |C_{-|n|}|^\sigma$ for $n\ge
0$. We further note that a model derivation of power law nonlinearity is 
given in ref.\cite{bik1}. When all points have the same nonlinearity 
strength, we have $\chi_n = 
\chi$, $n \in Z$. On the other hand for a dimeric nonlinear impurity, $\chi_n
= \chi (\delta_{n,0} + \delta_{n,1}), n \in Z$. Furthermore, the classical 
Hamiltonian which can generate Eq.(4) is
\begin{equation}
H = \sum_{n=-\infty}^{\infty} \frac{\chi_n}{\sigma + 2} |C_n|^{\sigma +2} + 
\frac{1}{2} \sum_{n=-\infty}^{\infty} (C_n C_{n+1}^\star + C_n^\star C_{n+1})
+ \frac{1}{2} V_0 ( C_0 C_1^\star + C_1 C_0^\star )
\end{equation}
where $V_0 = (\frac{1}{\sqrt K} - 1)$.

\section{A dimeric nonlinear impurity in the Cayley tree}

Here we are interested in the possible solutions for SL states due to 
a dimeric impurity. Hence we assume that 
\begin{eqnarray}
C_m &=& \phi_m e^{-iEt},\nonumber \\
\noindent & & {\rm where} \nonumber \\
\phi_m &=& \left ({\rm
sgn}(E) \eta \right )^{m-1} \phi_1 ~~{\rm for} ~m\ge 1 \nonumber \\
\noindent & &{\rm and}\nonumber \\
\phi_{-|m|} &=& \left (
{\rm sgn}(E) \eta \right )^{|m|} \phi_0 ~~{\rm for}~ m \le 0. 
\end{eqnarray}
Eq.(6) is the exact 
form of $\phi_m$ in the presence of a dimeric impurity and can be derived from
Green's function analysis. $\eta \in [0,1]$ is given by $\eta = \left ( |E|
- \sqrt{E^2-4}\right )/2$. Sgn($E$) denotes the signature of $E$. We further 
define $\beta = \phi_0/\phi_1$ if $|\phi_1| \le |\phi_0|$. Otherwise we invert 
the definition of $\beta$. Because of the symmetry in the system we will
obtain the same result. So, apparently $\beta \in [-1,1]$. However, we shall 
show later that for $\chi > 0$, negative values of $ \beta$ except $\beta = 
-1$ are not permissible. The analytical argument showing the impossibility 
of such a situation is presented in ref.\cite{bik1}. Of course, negative values 
of $\beta \ge -1$ will produce SL states in the antisymmetric set if $\chi < 
0$.   So, for $\chi > 0$, $\beta \in [0,1]$ and for $\chi < 0$, $\beta \in 
[-1,0]$. This assertion will also be substantiated here in due course. Now, 
from the normalization condition we get
\begin{equation}
|\phi_0|^2 = \frac{1 - \eta^2}{1 + \beta^2}
\end{equation}
Introducing Eqs.(6) and (7), and the definition of $\beta$ in the Hamiltonian 
(Eq.(5)) we get an effective Hamiltonian, $H_{eff}$ where
\begin{equation}
H_{eff} = {\rm sgn}(E) \eta + \frac{V \beta (1 - \eta^2)}{1 + \beta^2} 
+ \frac{
\chi}{\sigma +2} \frac{(1 - \eta^2)^{\sigma /2+1} (1 + |\beta|^{\sigma+2})}{
(1 + \beta^2)^{\sigma /2+1}}
\end{equation}
and $V=\frac{1}{\sqrt K}$ for the Cayley tree otherwise it is a parameter. 
The Hamiltonian consists of two variables, namely, $\beta$
and $\eta$ and two constants, namely, $\chi$ and $\sigma$. The stationary 
localized states correspond to fixed points of the reduced dynamical system
described by $H_{eff}$. 

\begin{center}
{\bf 1.}~~$|\beta| = 1$
\end{center}

We here consider the case where $|\beta| = 1$. 
For $\chi > 0$, $\beta = 1$ yields the symmetric set, while for $\beta = 
-1$ we get the antisymmetric set. In this limit the relevant equation governing the formation of SL states is obtained by setting $dH_{eff}/d\eta$ = 0. This in turn yields
\begin{equation}
\frac{1}{\chi} = 2^{-\sigma /2} \frac{\eta (1-\eta^2)^{\sigma /2}}{{\rm sgn}(E)
- {\rm sgn}(\beta )V\eta} = F(\sigma , \eta )
\end{equation} 

\noindent {\bf Case 1: {\bf $\sigma = 0$}}

\noindent Here we have a linear dimeric impurity in a Cayley tree. 
Since we are considering $\chi > 0$, for the symmetric case we must have 
sgn($E$) = 1. On the other hand for antisymmetric case sgn($E$) can take both 
the signs. If $V<1$, for the symmetric case $F(0,\eta )$ has a divergence at 
$\eta = \frac{1}{V}>1$. Since $\eta \in [0,1]$, this divergence occurs beyond 
the permissible range of $\eta$. However, in the permissible range of $\eta$,
$\frac{dF}{d\eta}=(1 - V\eta )^{-2} >0$. So, $F(0,\eta )$ is a monotonically
increasing function of $\eta$ and it assumes the permissible maximum value at
$\eta = 1$. This in turn then yields $\chi_{cr}=(1-V)$ and a SL state will
be obtained if $\chi \ge \chi_{cr}$. For the dimer in a Cayley tree we then 
need $\widetilde \chi \ge \widetilde \chi_{cr} = \sqrt{K}-1$. On the other 
hand for $V>1$, the divergence at $\eta = \frac{1}{V}$ is in the permissible 
range of $\eta$. So, even with an infinitesimally small value of $\chi$, we 
shall obtain a SL state. However, for $\chi > 0$, $\eta \in [0,\frac{1}{V}]$. 

We consider now the antisymmetric case with ${\rm sgn}(E) = +1$. It is easy to 
see that $F(0,\eta)$ is a monotonically increasing function of $\eta$. So, 
$F(0,\eta)$ takes the maximum possible value at $\eta =1$. This in turn 
gives $\chi_{cr}=1+V$. $V$ is implicitly assumed to be positive. So, we shall 
get a SL state for the Cayley tree if $\widetilde \chi \ge \widetilde \chi_{cr}
=\sqrt {K}+1$. For ${\rm sgn}(E)=-1$, $F(0,\eta)$ will diverge at $\eta=\frac{1}{V}$.  So, if $V<1$, we shall not obtain any SL state below the band of the 
host system. On the other hand if $V>1$, $F(0,\eta)$ diverges in the 
permissible range of $\eta$. However, we also have $F(0,1)=\frac{1}{V-1}$. 
Thus, $\chi_{cr} =V-1$ and a SL state below the band will be obtained if $\chi 
\le \chi_{cr}= V-1$. We now summarize  our findings on the linear dimer.

\noindent (i) $V<1$. No SL state will be obtained if $\chi < 1-V$. There is 
one SL state for $(1-V)<\chi <(1+V)$. But there are two SL states if $\chi > 
(1 + V)$\cite{econ}.
\noindent (ii) $V>1$. If $0<\chi \le (V-1)$, there are two SL states. One 
appears above the band and the other lies below the band. If $(V-1)< \chi <
(1+V)$, we have one SL state above the band. For $\chi \ge (V+1)$, we get 
two SL states and both appear above the band.

\noindent {\bf Case II}: $\sigma \ne 0$. 

Here we consider two cases, namely, 
$V<1$ and $V>1$ separately. 

\noindent{\bf (A)} $V<1$. Since $\chi$ is taken to be positive, 
in Eq.(9) we need 
${\rm sgn}(E)={\rm sgn}(\beta)=+1$. Again the divergence of $F(\sigma
, \eta )$ at $\eta =\frac{1}{V}$ is of no consequence. 
Furthermore, $F(\sigma, \eta )$ has at least one maximum at $\eta_m 
\in [0,1]$. In fact, $F(\sigma , \beta )$ has only one maximum. So,
there will be a $\chi_{cr}^s$ so that for $\chi > \chi_{cr}^s$ we shall
obtain two SL states and for $\chi < \chi_{cr}^s$, there will be no SL
state. On the other hand in the antisymmetric case we have ${\rm sgn}
(\beta )=-1$ but ${\rm sgn}(E)$ can be either +1 or -1. In the first
case (sgn($E$)=+1), $F(\sigma ,\eta )$ has no divergence but $F(\sigma,0)$ 
= 0 = $F(\sigma ,1)$.  So, $F(\sigma , \eta )$ has at least one (actually
one) maximum at $\eta_m^\prime \in [0,1]$. Consequently, we shall get another 
critical value of $\chi$, say $\chi_{cr}^a$ so that if $\chi > \chi_{cr}^a$ 
we shall obtain two SL states. For $\chi < \chi_{cr}^a$ there will be no 
SL state. We further note that $\chi_{cr}^a > \chi_{cr}^s$. In the second 
case (sgn($E$) = -1), $F(\sigma ,\eta )$ diverges at $\eta = \frac{1}{V} >1$. 
Furthermore, $F(\sigma , \eta )$ should be positive. Hence, $\eta \ge 
\frac{1}{V}$. Since allowed
values of $\eta \in [0,1]$, no SL state will be obtained below the
host band. So, we shall get three regions having no, two and four SL
states. Furthermore, in half of the states $\eta \rightarrow 1$ as
$\chi \rightarrow \infty$. So, these are unstable states. Equations
for critical lines in the $( \chi ,\sigma )$ plane separating three
regions are given in Appendix II.

\noindent {\bf(B)} $V>1$. In the symmetric case $F(\sigma ,\eta )$ 
diverges at $\eta
=\frac{1}{V} \le 1$ and $F(\sigma ,0) = 0 = F(\sigma ,1)$. Furthermore,  
$F(\sigma ,\eta ) \ge 0$ for $\eta < \frac{1}{V}$ and $F(\sigma ,\eta
) \le 0$ for $\eta > \frac{1}{V}$. Hence, a SL state will always be
obtained for $\chi >0$. The maximum value of $\eta$ the SL state can
take is $\frac{1}{V}$ and this will happen if $\chi \sim 0$. In the
antisymmetric case when ${\rm sgn}(E) = +1$, $F(\sigma ,\eta )$ has no
divergence for $\eta \in [0,1]$. Since, $F(\sigma ,0)$ = 0 = $F(\sigma 
,1)$, $F(\sigma ,\eta )$ has a maximum at $\eta_m^{''} \in [0,1]$. So, there 
will be a critical value of $\chi$ such that $\chi < \chi_{cr}$ no SL 
state will be obtained. On the other hand for $\chi > \chi_{cr}$, we shall 
have two SL states and in one of these states $\eta \rightarrow 1$ as
$\chi \rightarrow \infty$. So, one of the states is an unstable SL state.
Equation for the critical line in $(\chi ,\sigma )$ plane separating 
these two regions is also given in Appendix II. For ${\rm sgn}(E)=-1$,
$F(\sigma ,\eta )$ again diverges at $\eta = \frac{1}{V}$. However, for 
$F(\sigma, \eta )$ to be positive we need $\frac{1}{V} \le \eta \le 1$.
Since $F(\sigma ,1)$ = 0, we shall always get a SL state irrespective
of the value of $\chi$. Furthermore, the SL state will appear below the
host band. The minimum value of $\eta$ that a SL state in this case can
attain is $\frac{1}{V}$. This will happen if $\chi \sim 0$ and $\eta 
\rightarrow 1$ as $\chi \rightarrow \infty$. So, this is an unstable SL
state. Finally for $V>1$, we have two regions. The region below the 
critical line has two SL states and that above the critical line has 
four SL states. As usual, half of these states are unstable.

\begin{center}
{\bf 2.}~~ $|\beta |\ne 1$ 
\end{center}

We consider here the scenario, $|\beta | \ne 1$. $H_{eff}$ (Eq.(8)) has
now two dynamical variables, $\eta$ and $\beta$. So, the relevant equation 
governing the formation of SL states is obtained by setting 
$\partial{H_{eff}}/\partial{X_i}$ = 0, where $X_1 = \eta$ and 
$X_2 = \beta$. We note in passing that for $\sigma = 0$, $\partial H_{eff}/
\partial \beta$ = 0 yields $\beta =\pm 1$. From the first condition 
($\partial H_{eff}/\partial \eta = 0$) we obtain
\begin{equation}
{\rm sgn}(E) \eta = \frac{{\rm sgn}(\beta)}{V} \frac{|\beta|^{-\sigma /2}
- |\beta|^{\sigma /2}}{|\beta|^{-(\sigma /2 + 1)} - |\beta|^{\sigma /2 + 1}}
\end{equation}
So, $\eta$ is a symmetric function of $\beta$ and $\beta^{-1}$ as enunciated 
earlier. Furthermore, since $\eta \ge 0$, if $V > 0$, sgn($E$)=sgn($\beta$).
In subsequent discussion we assume that sgn($\beta$) =+1. When $|\beta| 
\rightarrow 1$, from Eq.(10) we obtain,
\begin{equation}
\eta_u = \frac{1}{V} \frac{\sigma}{\sigma +2}
\end{equation}
The second condition ($\partial H_{eff}/\partial \beta$ = 0) yields
\begin{equation}
\frac{1}{\chi} = \frac{{\rm sgn}(\beta )|\beta|\left (1-|\beta|^\sigma \right 
)}{V\left (1+\beta^2\right )^{\sigma /2}\left (1 - \beta^2\right )} \left (1
-\eta^2\right )^{\sigma /2}
\end{equation}
Since $(1-\eta^2)$ for $\eta \in [0,1]$ is a positive semidefinite quantity,
for $V>0$, $\chi$ and $\beta$ will possess the same sign. In other words, SL
states with $|\beta| \ne 1$ in the antisymmetric set are not possible. It is
trivially seen that the right hand side of Eq.(12) is also a symmetric function
of $\beta$ and $\beta^{-1}$. We further note that introduction of Eq.(10) in 
Eq.(12) makes the right hand side of Eq.(12) an explicit function of $\sigma$ 
and $\beta$. We call this function $F(\sigma ,\beta)$.

Since $\sigma =2$ is physically more relevant we consider this case in detail.
In this situation we have
\begin{equation}
\frac{1}{\chi} = \eta (1 - \eta^2) = F(2, \eta) = g(\eta)
\end{equation}
We note that $g(\eta)$ has one and only one maximum at $\eta_m^2 = \frac{1}
{3}$ for $\eta \in [0,1]$. Furthermore, $\eta_u = (2V)^{-1}$. So, if $V<0.5$ 
or $K>4$, $\eta_u > 1$. Since $g(\eta)$ in this case is defined for $\eta 
\in [0,1]$, the maximum of $g(\eta)$ lies in the permissible range of $\eta$.
On the other hand if $V \ge 0.5$ ($K \le 4$), $\eta_u \le 1$. So, $\eta \in
[0,\eta_u]$. Then for $\eta_m$ to stay in the allowed range of $\eta$, we need 
$\eta_m^2 \le \eta_u^2$. This in turn yields $V \le 0.8660$ or $K \ge 1.33$.
So, there will be a lower critical value of $\chi$, $\chi_{crl}$ such that 
$\chi < \chi_{crl}$ there will be no SL state and for $\chi > \chi_{crl}$ there
can be two SL states. From Eq.(13) we further obtain $\chi_{crl} = \widetilde 
\chi_{crl}
/\sqrt{K}$ = 2.5980. Again for $V \le 0.5$, since $\eta \in [0,1]$ and $g(0)$
= 0 = $g(1)$, we shall obtain two SL states for $V \le 0.5$ and $\chi > 
2.5980$. In one of the states $\eta \rightarrow 1$ as $\chi \rightarrow 
\infty$. So, one state 
is an unstable state. On the other hand, $0.5 \le V \le 0.8660$, 
we get $\chi_{cru}=
\frac{8V^3}{4V^2-1}$ and $\chi > \chi_{cru}$, we get only one SL state. Then,
if $V \le 0.8660$ and $\chi_{crl} < \chi \le \chi_{cru}$ we have two SL states. 
In one state $\eta \rightarrow \eta_u$ as $\chi \rightarrow \infty$. So, one
state is unstable.   Thirdly, for $V > 0.8660$, $\eta \in [0,\eta_u]$ and 
$\eta_u < \eta_m$. So, $g(\eta)$ takes the maximum value at $\eta_u$ and the 
corresponding critical value of $ \chi$ is $\chi_{cru}$. We note that for 
$V=1$, $\chi_{cru} =8/3$ \cite{bik1,bik2} and for $V=\sqrt 2$, $\chi_{cru} 
=3.232$. For $\chi 
> \chi_{cru}$ we shall obtain one stable SL state. 

We now consider the general $\sigma$. Substituting $\eta_u =1$ in Eq.(11)
we obtain $\sigma^\prime = 2V/(1-V)$ if $V<1$. For the Cayley tree it 
translates to $\sigma^\prime = 2/(\sqrt K -1)$. For simplicity we break 
the discussion in two cases.

\noindent{{\bf Case I: $\sigma \ge \sigma^\prime$}} 

\noindent Here $\eta_u \ge 1$. So, $\eta \in [0,1]$ but 
$\beta \in [0,\beta_u]$ where $\beta_u \le 1$. Furthermore, $F(\beta_u,\sigma)
$ = 0 = $F(0,\sigma)$. So, for a given $\sigma \ge \sigma^\prime$, $F(\beta ,
\sigma )$ has at least one maximum at $\beta_m \in [0,\beta_u]$. For $K=4$ or 
$V=0.5$, $\sigma^\prime = 2$. In Fig.(2) we have plotted  $F(\beta,\sigma)$ 
for $\sigma = 2.5$. We see that there is only one maximum. Due to the maximum
at $\beta_m$, in the $(\chi,\sigma)$ plane there will be a critical line 
separating the no state region from the region containing two SL states. 
The equation of the critical line is $\chi_{cr}^{(1)}$ = $[F(\beta_m,\sigma
)]^{-1}$ where $\beta_m$ is the solution of $\partial F / \partial \beta$=0.
Since in one of the states $\eta \rightarrow 1$ as $\chi \rightarrow \infty$,
the two states region has one unstable state. Furthermore, as we go along the 
critical line $\beta \rightarrow 0$, $\eta \rightarrow 0$ and $\chi 
\rightarrow \infty$. This implies that in the stable state, the amplitude 
gets preferentially localized in one of the dimer sites as $\sigma \rightarrow 
\infty$. $\chi \rightarrow \infty$ because in the limit a monomer localized 
state is formed. 

\noindent{\bf Case II: $\sigma < \sigma^\prime$}
 
Since $\eta_u < 1$, $\beta \in [0,1]$. Consequently,
there will be a critical value of $\chi$, $\chi_{cr}^{(2)}$ given by 
\begin{equation}
\chi_{cr}^{(2)} = \frac{2V}{\sigma} \left (\frac{1-\eta_u^2}{2}\right )^
{-\sigma /2}.
\end{equation}
For $\chi > \chi_{cr}^{(2)}$, we shall obtain at least one SL state. Since $
\chi_{cr}^{(2)} \rightarrow \infty$ as $\sigma \rightarrow 0$ and $\sigma 
\rightarrow \sigma^\prime$ ($\eta_u \rightarrow$ 1), $\chi_{cr}^{(2)}$ will 
assume a minimum value at $\sigma_{min}$. $\sigma_{min}$ is obtained from 
$d\chi_{cr}^{(2)} / d\sigma$ =0. However, it gives a very complicated algebraic 
equation in $\sigma$. When $\sigma \rightarrow 0$, $\eta \rightarrow 0$.
So, we obtain a SL state localized mostly on the dimer. Precisely for this 
$\chi_{cr}^{(2)} \rightarrow \infty$ as $\sigma \rightarrow 0$. Again $\sigma$ 
increases, $\eta$ increases. This will require lower values of $\chi$.  So, 
for $0 \le \sigma \le \sigma_{min}$ and $\chi \ge \chi_{cr}^{(2)}$, the system 
yields a SL state with $\beta \ne 1$. On the other hand, for $\sigma_{min} 
\le \sigma \le \sigma^\prime$, albeit $\eta$ increases towards unity as 
$\sigma \rightarrow \sigma^\prime$, $\chi_{cr}^{(2)}$ increases towards 
infinity.   So, the SL state obtained for $\sigma > \sigma_{min}$ (in fact
$\sigma_{cr}$ defined later) and $\chi 
\ge \chi_{cr} ^{(2)}$ is unstable.

$F(\beta,\sigma)$ also develops a local maximum at $\beta_m <1$. This is
shown in Fig.(2) for $K=4$ ($V$=0.5) and $\sigma = 1.5$. So, there will be 
a critical value of $\sigma$, $\sigma_{cr}$ defined as follows. If $\sigma
=\sigma_{cr} + \delta$ and $\delta \rightarrow 0$, there exists an $\epsilon 
\rightarrow 0$ depending on $\delta$ so that $\beta_m=1-\epsilon$. So, for 
$\sigma \ge \sigma_{cr}$ there will be a lower critical value of $\chi$,
$\chi_{cr}^{(3)}$ and $\chi < \chi_{cr}^{(3)}$ no SL state will be obtained.
It is trivially seen that $\chi_{cr}^{(3)} = [F(\beta_m,\sigma)]^{-1}$. For
a given $\sigma > \sigma_{cr}$ the upper critical value is $\chi_{cr}^{(2)}(
\sigma)$. Then in the $(\chi,\sigma)$ plane we have a V-shaped region with
boundaries $\sigma_{cr} \le \sigma \le \sigma^\prime$ and $\chi > \chi_{cr}
^{(3)}(\sigma_{cr})$ but $\chi^{(2)}_{cr}(\sigma) > \chi > \chi^{(3)}_
{cr}(\sigma)
$. This region contains two SL states with $\beta \ne 1$. 
However one of the states is unstable because $\eta \rightarrow \eta_u$
as $\chi \rightarrow \infty$. We further note that the lower boundary of
the V-shaped region joins smoothly with the critical line for $\sigma \ge
\sigma^\prime$. 

\noindent{\bf Case III: {\bf $V>1$}} 

The scenario is very similar to the $V=1$ case discussed in ref.\cite{bik1}. 
However $\sigma_{cr}$ moves to a higher value depending on the magnitude of
$V$.

The total phase diagram of SL states for $V=0.5$ ($K=4$) is shown in Fig.(3). 
The number of possible SL states in each region is indicated. We 
again note that the maximum number of SL states is six. The total phase 
diagram for $V=\sqrt 2$ is shown in Fig.(4). The maximum number of SL states 
is found to be five. Furthermore, the $\sigma =2$ line has no special 
significance. The stability diagram \cite{prl} of SL states for dimeric 
nonlinear impurity in a Cayley tree is shown in the $(\eta,\sigma)$ 
plane in Fig.(5).

\section{all sites nonlinear}

\noindent {\bf 1.~~Inter site peaked and dipped solutions}

We consider here the formation of SL states in a Cayley tree with each site
having a power law nonlinearity and the same coupling constant, $\chi$.
But due to the transformation proposed here, the problem reduces to the 
study of 
formation of SL states in a one dimensional system with a bond defect between 
the zeroth and the first site. The tunneling matrix between the sites, $V$
is reduced by a factor of $1/\sqrt K$ and site energies are 
\begin{eqnarray}
\epsilon_n &=& V^{n-1} \chi |C_n|^\sigma ~~~~~{\rm if}~~ n \ge 1 \nonumber \\
& &{\rm and} \nonumber \\
\epsilon_{-|n|}&=& V^{|n|} \chi |C_{-|n|}^\sigma ~~~~~{\rm if}~~n \ge 0.  
\end{eqnarray}
Since we have a bond defect, from the symmetry consideration, we use the dimer 
ansatz first to study the problem. The application of the method outlined 
here and in
other places \cite{bik1,bik2} yields the equation governing the formation of SL states
in this system. It is given by
\begin{equation}
\frac{2^{\sigma /2}}{\chi} = \frac{\eta (1-\eta^2)^{\sigma /2} \left (1 - 
(V\eta)^\sigma \right )}{\left ({\rm sgn}(E)-{\rm sgn}(\beta)V\eta\right )
\left (1 - \eta^2 (V\eta)^\sigma\right )} = G_(\eta,\sigma)
\end{equation}
$\eta$ is defined earlier \cite{bik3}. We are also assuming that $\chi >0$. 
Since $[1-(V\eta)^\sigma] \rightarrow -\sigma~ln(V\eta)$ as $\sigma 
\rightarrow 0$, $G(\eta,\sigma) \rightarrow 0$ as $\sigma \rightarrow 0$. 
Consequently $\chi \rightarrow \infty$. This implies that no SL state will 
be formed in this limit.   We now consider various cases.  

\noindent (I) {\bf Symmetric Case}

Here sgn($E$) = sgn($\beta$) = +1. We note that $G(\eta,\sigma)$ have a 
removable singularity and a divergence  at $\eta_0 = 1/V$ and $\eta_1 = 
1/V^{\frac{\sigma}{\sigma+2}}$ respectively. But for $V<1$, $\eta_1>1$. 
So, the singularity of $G(\eta,\sigma)$ at $\eta_1$ will
not play any role in the formation of SL states. Again $G(0,\sigma)$ = 0 = 
$G(1,\sigma)$. So, $G(\eta,\sigma)$ will have at least one maximum at 
$\eta_m \in [0,1]$. It can be seen numerically that $G(\eta,\sigma)$ has 
only one maximum for $\eta \in [0,1]$. Consequently in the $(\chi,\sigma)$
plane there will be a critical line separating two states region from the 
no state region. Since one of the states in the two states region $\eta 
\rightarrow 1$ as $\chi \rightarrow \infty$, it is an unstable state.
On the other hand for $V>1$, $G(\eta,\sigma)$ has a divergence at $\eta_1<1$.
So, in this case the system will always produce a SL state even if $\chi$ 
is infinitesimally small. Furthermore, $lim_{\epsilon\rightarrow 0}G(\eta_1-
\epsilon,\sigma)\rightarrow \infty$ from the positive side only. So, for $\chi 
>0$, $\eta_{max}=\eta_1$ and $\eta \rightarrow 0$ as $\chi \rightarrow 
\infty$. Hence this is a stable SL state.

\noindent (II) {\bf Antisymmetric Case}

In this limit sgn($\beta$) = -1 but sgn($E$) can be either +1 or -1. We note 
that for sgn($E$)=+1, $G(\eta_0,\sigma)$=0 here. But for $V<1$ both $\eta_0$ 
and $\eta_1$ lie beyond unity. Since $G(\eta,\sigma)$=0 both at $\eta$=0 and 
$\eta$=1, it has a maximum at $\eta_m \in [0,1]$. This implies that in the $
(\chi,\sigma)$ plane there will be a critical line. This line again separates 
the no state region and the two states region. Furthermore, in the two state 
region one state will be unstable for the same argument given earlier. 

For $V>1$, we note that $G(\eta,\sigma)$ goes to zero and infinity at $\eta_0$ 
and $\eta_1$ respectively. Furthermore, if $\sigma$ is finite, $\eta_0 < 
\eta_1$. So, for $\eta \in (\eta_0,\eta_1)$, $G(\eta,\sigma)$ is negative.
This in turn implies that $lim_{\epsilon\rightarrow 0}G(\eta_1-\epsilon,\sigma)
\rightarrow -\infty$ and $lim_{\epsilon\rightarrow 0}G(\eta_1+\epsilon,\sigma)
\rightarrow \infty$. Consequently, $lim_{\epsilon\rightarrow 0}G(1-\epsilon,
\sigma) \rightarrow 0$ from the positive direction for $\eta \in [\eta_1,1]$.
Therefore, we shall always obtain a SL state even if $\chi$ is infinitesimally 
small. In this SL state, however, $\eta_{min}=\eta_1$ and as $\eta \rightarrow 
1$, $\chi \rightarrow \infty$. So, this is an unstable state. We further note 
that $G(0,\sigma)$=0=$G(\eta_0,\sigma)$. Then there will be a maximum of $G(
\eta,\sigma)$ at $\eta_m \in [0,\eta_0]$. So, there will be a critical line in 
the $(\chi,\sigma)$ plane also. This line will separate one state region and 
three states region. It is further seen that two of the states in the later 
region are unstable. We can also have sgn($E$)=-1. However, for $V<1$ no SL
states will be obtained in this limit. On the other hand for $V>1$, $lim_{
\epsilon \rightarrow 0}G(\eta_1\mp\epsilon,\sigma)\rightarrow\mp\infty$. So,
we shall always get a SL state below the band even if $\chi$ is infinitesimally
small. However, this SL state is unstable.

We now combine our results to obtain the phase diagram. For $V<1$, we have 
three regions, namely I, II and III containing no SL state, two SL states and 
four SL states respectively. This is shown in Fig.(6) for $K=4$ or $V=0.5$.
The stability diagram \cite{prl} of SL states for this case is shown in 
Fig.(7). However,
for $V>1$ we do not have no SL state region. Instead we have a three state 
region separated  from a five state region by a critical line. One of these 
states appears below the band. In the three state region we have two unstable 
states while in the other region we have three unstable states. For $V=\sqrt 
2$, the phase diagram is shown in Fig.(8).

\noindent {\bf 2.~~On site peaked soliton}

We discuss here the formation of on-site soliton in the transformed system
described by the system given by the Eq.(2). For this purpose we first consider
a power law nonlinear impurity with strength, $\chi$ embedded at the zeroth 
site. After introducing the dimeric ansatz in the appropriate form of the 
Hamiltonian we obtain
\begin{equation}
H_{eff}=\frac{\chi}{\sigma+2}\left ( \frac{1-\eta^2}{1+\beta^2} \right )^{
\sigma/2+1}+{\rm sgn}(E)~\eta+V\beta\left(\frac{1-\eta^2}{1+\beta^2}\right)
\end{equation}
Again relevant equations are obtained by $\partial H_{eff}/\partial X_i$=0 
where $X_1=\beta$ and $X_2=\eta$. After a trite algebra we then obtain $\beta
={\rm sgn(E)} V\eta$ and 
\begin{equation}
\frac{{\rm sgn}(E)}{\chi}=\frac{\eta(1-\eta)^{\sigma/2}}{(1+V^2\eta^2)^
{\sigma/2} (1-V^2\eta^2)}=f(\eta,\sigma)
\end{equation}
For $V=1/\sqrt K$, Eq.(18) describes the formation of SL states due to a 
nonlinear impurity in a Cayley tree. This has been discussed in detail in 
ref.\cite{bik1}. So, we see that the dimeric ansatz reduces to the appropriate 
monomeric ansatz. When $V>1$, $f(\eta,\sigma)$ has a divergence at $\eta_u=1/V$. 
Furthermore we have, $f(0,\sigma)=0=f(1,\sigma)$ and $\lim_{\epsilon
\rightarrow 0}f(\eta_u-\epsilon,\sigma)\rightarrow \infty$. So, we shall 
obtain two SL states even if $\chi$ is infinitesimally small and $\sigma>0$.
However, one state will appear below the band. In this state $\eta_{min}
=\eta_u$ and $\eta\rightarrow 1$ as $\chi\rightarrow\infty$. So, this is an 
unstable state. For, $\sigma=0$, this state will appear if $0<\chi<(V^2-1)$.

To study the formation of on-site peaked SL states in the fully nonlinear chain
we put $\beta ={\rm sgn}(E) V\eta$ in $H_{eff}$ given by Eq.(17). We 
then set $\partial H_{eff}/\partial \eta$ = 0. After a trite algebra we finally 
obtain 
\begin{equation}
\frac{1}{\chi}=\frac{\eta\left(1-\eta^2\right)^{\sigma/2}\left(1+V^{\sigma+2}
\eta^{\sigma+4}\right)} {\left(1+V^2\eta^2\right)^{\sigma/2}\left(1-V^\sigma
\eta^{\sigma+2}\right)^2}\frac{\left(1-V^\sigma\eta^\sigma\right)}{\left
(1-V^2\eta^2\right)}
=f_1(\eta,\sigma)
\end{equation}
When V=1, Eq.(19) reduces to the relevant equation in ref. \cite{bik3}. 
We note that 
$f_1(\eta,\sigma)$ has a removable singularity at $\eta_u$ and a divergence
at $\eta_1=V^{\frac{-\sigma}{\sigma+2}}$. However, for $V<1$ the divergence
at $\eta_1$ does not play any role in the formation of SL states. Since 
$f_1(0,\sigma)$=0=$f_1(1,\sigma)$ we expect at least one maximum of $f_1(\eta,
\sigma)$ at $\eta_m \in [0,1]$. It is seen numerically that for $\eta \in 
[0,1]$ $f_1(\eta,\sigma)$ has only one maximum. So in the $(\chi,\sigma)$ plane there is a critical line separating the no state region from the two states 
region. Of course, one of the states is unstable. In Fig.(6) the critical line
is shown by solid curve for $K$ = 4. On the other hand, for $V>1$, $f_1
(\eta,\sigma)$
diverges at $\eta_1$. Since, $lim_{\epsilon\rightarrow 0}f_1(\eta_1\mp\epsilon,
\sigma)\rightarrow \infty$ we have also two states and these states are formed 
even if $\chi$ is infinitesimally small. One of the states is unstable. However,
numerical calculation shows that there exists a critical value of $\sigma$ say
$\sigma_{cr}$ such that for $\sigma > \sigma_{cr}$ there will be a four state 
region bounded by two critical values of $\chi$. For example, for $V=\surd 2$
$\sigma_{cr} \sim 3.85$.
But the four states region occurs at larger value of $\chi$. Two of the states 
are again unstable.

\section{summary}

The DNLSE with general power law nonlinearity is used to study the formation 
of stationary localized states in the Cayley tree. Two cases, namely, a dimeric
nonlinear impurity and the fully nonlinear system are considered. To facilitate
the study a transformation is devised to map the system to an one dimensional
system with a bond defect. We also note in passing that the problem can also be
mapped to an half infinite chain with a bond defect between the zeroth site 
and the subsequent site. The Cayley tree Green's function for the problem can
be obtained from the transformed system. This is discussed in Appendix I.

The formation of SL states is studied by analyzing the fixed point equations
of the reduced dynamical system. This is obtained by introducing the dimeric 
ansatz in the appropriate Hamiltonian.
For the linear dimer our results agree with known results. In case of a 
nonlinear dimer impurity, the system is found to sustain two types of SL 
states and altogether a maximum of {\it six} types of SL states can be 
obtained. In one case the absolute amplitude at two sites are equal. In 
the second category we find states with unequal amplitudes. In this aspect 
our results are very similar to what we obtain for a one dimensional chain. 
There are, however, some differences. In the one dimensional chain the no 
state region is obtained for $\sigma \ge 2$. Furthermore, the $V$ region 
extends to infinity ($\sigma^\prime \rightarrow \infty$). Here, no state 
region is obtained from $\sigma = 0$, and $\sigma^\prime = \frac{2}{(\sqrt 
K -1)}$. So, the $V$ region shrinks as $K$ increases. For $\sigma = 2$, 
we find that for $K > 1.33$ ($V < 0.86$), a third critical value of $\chi$,
$\chi_{cr} = \widetilde \chi_{cr}/\sqrt K$ = 2.5980. This is a very 
interesting result. The corresponding $\chi_{cr}$ for $K=1$ is 8/3. The 
stability and the phase diagram of SL states are discussed in detail. 

The formation of SL states in the fully nonlinear Cayley 
tree is also considered. For the on site peaked solution, the appropriate 
ansatz is derived. In the perfect nonlinear chain, a three SL states region 
exists for the on site peaked SL state. For the Cayley tree, this region is 
absent. Instead for all cases, we find a two states region and a no state 
region separated by respective critical lines. Along with this the case where
$V$ (in relation to other hopping element) $> 1$ is considered. For this case
we show that under certain conditions, states can appear both below and above
the band. Furthermore, we also find a four states region.

\section{Appendix I}

We show here that a subset of amplitude Green's functions of Eq.(4) with 
$\epsilon_n = 0$, $n \in Z$ yields the amplitude Green's functions of a 
particle moving 
on a Cayley tree with connectivity, $K$. We first note that $G_{n,m}(\widetilde
E = E\sqrt K) = \frac{1}{\sqrt K} \widetilde C_m(0) G_{n,m}(E)$. Furthermore, 
from the transformations, we have $\widetilde C_m(0) = C_m(0)/K^{(m-1)/2}$ if 
$m \ge 1$ and $\widetilde C_m(0) = C_m(0)/K^{|m|/2}$ if $m \le 0$. The 
Hamiltonian $(H)$ that yields Eq.(4) is $H = H_0 + H_1$ where,
\begin{eqnarray}
H_0 & = & \sum_n \left ( a_n^\dagger a_{n+1} + a_{n+1}^\dagger a_n \right ) 
\nonumber \\
& &{\rm and} \nonumber \\
H_1 & = & \left (\frac{1}{\surd K} - 1) \right ) \left ( a_0^\dagger a_1 + 
a_1^\dagger a_0 \right )
\end{eqnarray} 
$a_n (a_n^\dagger)$ destroys (creates) a particle at the $n$-$th$ site. 
We define $V_0
= (1/\sqrt K - 1)$. We further have $G(E) = (E - H)^{-1}$ and $G_0(E) =
(E - H_0)^{-1}$. These two operators are related by $(I - G_0 H_1)G = G_0$
where $I$ is the identity operator. By defining $<n|G(E)|m> = G_{n,m}(E)$ and
with similar definition for $G_{0(n,m)}(E)$ we obtain from the relation 
between $G$ and $G_0$
\begin{equation}
G_{n,m}(E) + V_0 \left (G_{0(n,1)}(E) G_{0,m}(E) + G_{0(n,0)}(E) G_{1,m}(E)
\right ) = G_{0(n,m)}(E).
\end{equation}
We have then two unknowns, namely, $G_{0,m}(E)$ and $G_{1,m}(E)$. After some 
algebra
we obtain
\begin{equation}
\left ( 
\begin{array}{c}
{G_{0,m}(E)}\\
{G_{1,m}(E)}
\end{array}
\right )=\frac{1}{D}
\left (
\begin{array}{cc}
{1-V_0 G_{0(1,0)}(E)}&{V_0 G_{0(0,0)}(E)} \\
{V_0 G_{0(0,0)}(E)}& {1-V_0 G_{0(0,1)}(E)}
\end{array}
\right )
\times
\left (
\begin{array}{c}
{G_{0(0,m)}(E)}\\
{G_{0(1,m)}(E)}
\end{array}
\right )
\end{equation}

and
\begin{equation}
D=\left (1 - V_0 G_{0(0,1)}(E) \right ) \left (1 - V_0 G_{0(1,0)}(E) 
\right ) - V_0^2 G^2_{0(1,0)}(E)
\end{equation}
We again note that
$G_{0(m,n)}(E)=[{\rm sgn}(E)]^{n-m+1} G_{0(0,0)}(|E|) \eta^{|n-m|}$ 
and $G_{0(0,0)}(|E|)=1/\sqrt{E^2-4}$ for $|E| > 2$. Hence,
$G_{0(0,0)}(|E|)=(1-\eta^2)/\eta$. After some simple algebra we obtain
$D=[(K-1)\eta G_{0(0,0)}(|E|) + K]/K$. Furthermore, from Eq.(22) and 
relevant transformations we obtain
\begin{equation}
\widetilde G_{0,0}(\widetilde E)=\frac{1}{\sqrt K}G_{0,0}(E)
={\rm sgn}
(\widetilde E)
\frac{2K}{(K-1)|\widetilde E|+(K+1)\sqrt{\widetilde E^2-4K}}
\end{equation}

Again from Eq.(22) we can easily show that for $m > 0$
\begin{equation}
\widetilde G_{o,m}(\widetilde E) = \frac{1}{\sqrt K} \frac{1}{K^{(m-1)/2}} 
G_{0,m}(E)
= [{\rm sgn}(\widetilde E)]^{m+1} \widetilde G_{0,0}(|\widetilde E|) 
\left (\frac{2}{|\widetilde E| + \sqrt{\widetilde E^2 - 4 K}}\right )^{|m|}
\end{equation}
when $\widetilde G_{0,0}(|\widetilde E|)$ is given by Eq.(24). On the 
other hand for $m<0$, we have
\begin{equation}
\widetilde G_{0,-|m|}(\widetilde E) = \frac{1}{\sqrt K} \frac{1}{K^{|m|/2}}
G_{0,-|m|}(E) = \widetilde G_{0,m}(\widetilde E)
\end{equation}
From Eq.(22) we further obtain, for $m \ge 1$
\begin{equation}
\widetilde G_{1,m}(\widetilde E) 
= \frac{1}{\sqrt K} \frac{1}{K^{(m-1)/2}}
G_{1,m}(E) 
= [{\rm sgn(\widetilde E)}]^m \widetilde G_{0,0}(\widetilde E)\left ( 
\frac{2}{|\widetilde E| + \sqrt{\widetilde E^2 - 4K}}
\right )^{m-1} 
= \widetilde G_{0,m-1}(\widetilde E)
\end{equation}
On the other hand for $m \le 0$ we obtain 
\begin{equation}
\widetilde G_{1,-|m|}(\widetilde E) = \frac{1}{\sqrt K} \frac{1}{K^{|m|/2}}
G_{1,-|m|}(E) 
= \widetilde G_{0,|m|+1}(\widetilde E)
\end{equation}
Since in relation to the Cayley tree we are dealing with a translationally
invariant problem, the choice of origin is arbitrary. So, (0,1) bond can be 
rechristened ($n,n+1$) without any loss of generality. If the shift in 
the origin is 
incorporated in the transformation, $G_{0,m}(E)$ and $G_{1,m}(E)$ will be 
transformed to $\widetilde G_{n,m}(\widetilde E)$ and $\widetilde G_{n+1,m}
(\widetilde E)$ respectively. Consequently our result will agree in full 
with the calculation in ref.\cite{econ}.

\section{Appendix II}
We derive here the equation of the critical line for the symmetric state
with $\beta$=1 and for the antisymmetric state. Note that $\chi >0$ and 
$V>0$. The equation to be considered for the purpose is 
\begin{equation}
\frac{2^{\sigma /2}}{\chi}=\frac{\eta (1-\eta^2)^{\sigma /2}}{{\rm sgn}(E)
- {\rm sgn}(\beta)V\eta} = F(\eta,\sigma)
\end{equation}
Then the equation of the critical line in $(\chi,\sigma)$ plane is 
\begin{equation}
\chi^\pm_{cr} = \frac{2^{\sigma/2}}{F(\eta^\pm_{cr},\sigma)}
\end{equation}
In Eq.(30) $\pm$ refers to the symmetric and the antisymmetric cases 
respectively. To find $\eta_{cr}^\pm$ we set $\partial F / \partial \eta$ =
0. This in turn yields 
\begin{equation}
\sigma V\eta^3\mp (\sigma+1)\eta^2\pm 1 =0
\end{equation}
In Eq.(31) the upper sign refers to the symmetric case and the lower 
sign to the antisymmetric case. For $V=1$, we find that 
\begin{equation}
\sigma \eta^2 \mp \eta -1 =0
\end{equation}
Eq.(32) has been derived in ref.\cite{bik1}. $\eta_{cr}$ is a real positive 
root of Eq.(31) and it must be less than unity. The expression for the 
$\eta_{cr}$ is found to be 
\begin{equation}
\eta_{cr}^{\pm} = \frac{\sigma +1}{3\sigma V} \pm Im \left (\frac{\sqrt 3}
{6\sigma V} \left (\frac{Q}{\sqrt[3]{2}} - \frac{\sqrt[3]{2} \left 
(\sigma+1 \right )^2}{Q} \right ) \right)
\mp \frac{1} {6\sigma V}
\left (\frac{\sqrt[3]{2} \left (\sigma +1 \right )^2}{Q} + \frac{Q}
{\sqrt[3]{2}} \right )
\end{equation}
where, $Q$=$\sqrt[3]{(A+3 \sigma V \sqrt{3B})}$, $A$=$2$+$6 \sigma
$+$6 \sigma^2$+$2 \sigma^3$-$27 \sigma^2 V^2$ and $B$=-$4$-$12 \sigma$-
$12 \sigma^2$-$4 \sigma^3$+$27 \sigma^2 V^2$. In Eq.(33) $Im$ refers to the 
imaginary part. While the upper sign refers to the symmetric case, lower
sign is for the antisymmetric case.

\begin{figure}
\caption{Cayley tree with connectivity 2. All bonds are of same length.}
\end{figure}
\begin{figure}
\caption{$F(\beta,\sigma)$ as a function of $\beta$ for $\beta \ne 1$ in 
case of a dimeric nonlinear impurity in a Cayley tree. Here $K$=4.
Solid, dotted and dashes curves are for $\sigma$ = 1.5, 2 and 2.5 
respectively}
\end{figure}
\begin{figure}
\caption{Total phase diagram of SL states of a Cayley tree in presence of 
a dimeric nonlinear impurity. Here $K=4 (V=0.5)$. $\chi^s$ and $\chi^a$ 
represents the critical lines for symmetric and antisymmetric case 
respectively. 
$\chi_{cru}$ and $\chi_{crl}$ represents the upper and lower critical line
respectively for $\beta \ne 1$ case. $\sigma_{min}$, $\sigma_{cr}$ and
$\sigma^\prime$ are 
shown. Numbers indicate the number of possible SL states in those regions
in the $(\chi,\sigma)$ plane}
\end{figure}

\begin{figure}
\caption{Total phase diagram for SL states of a one dimensional chain with 
a dimeric nonlinear impurity and a bond defect in between the impurity 
sites. Here $V=\surd{2}$. Numbers indicate the number of SL states in those 
regions. The unnumbered closed triangular small region contains four SL 
states.}
\end{figure}

\begin{figure}
\caption{Stability diagram for SL states in a Cayley tree with a dimeric
nonlinear impurity. Here $K=4 (V=0.5)$. Stability of the states in various 
regions are marked in the figure.}
\end{figure}

\begin{figure}
\caption{Phase diagram for SL states in a fully nonlinear Cayley tree. The 
solid line is the critical line for the on site (zeroth site) peaked solution. 
The lower dotted line is the critical line for the inter site peaked 
(symmetric) solution and the uppermost line defines the critical line for 
the inter site dipped (antisymmetric) solution. Here $K=4 (V=0.5)$. For inter
site solutions region I, II and III contains no, two and three SL states
respectively. For on site solutions there is no state below the solid 
curve and two states above the solid curve.}
\end{figure}

\begin{figure}
\caption{Stability diagram for SL states in a fully nonlinear Cayley tree. 
Here $K=4 (V=0.5)$. Stability of the states in various regions are marked 
in the figure.}
\end{figure}

\begin{figure}
\caption{Total phase diagram for SL states of a fully nonlinear one 
dimensional chain a dimeric nonlinear impurity and a bond defect in 
between the impurity sites. Here $V=\surd 2$. The region I contains 
three SL states and the region II contains five SL states.}
\end{figure}
\end{document}